# Efficient Medium Access Control for Low-Latency Industrial M2M Communications

Anwar Ahmed Khan* [1,2], Indrakshi Dey[2]

[1]*Millennium Institute of Technology & Entrepreneurship,* Karachi, Pakistan
email: anwar.ahmed@mite.edu.pk
0000-0002-2237-5124

[2] *Walton Institute for Information and Communication Systems Science*
South East Technological University, Waterford, Ireland
email: indrakshi.dey@waltoninstitute.ie
0000-0001-9669-6417

***Abstract*** -Efficient medium access control (MAC) is critical for enabling low-latency and reliable communication in industrial Machine-to-Machine (M2M) networks, where timely data delivery is essential for seamless operation. The presence of multi-priority data in high-risk industrial environments further adds to the challenges. The development of tens of MAC schemes over the past decade often makes it a tough choice to deploy the most efficient solution. Therefore, a comprehensive cross-comparison of major MAC protocols across a range of performance parameters appears necessary to gain deeper insights into their relative strengths and limitations. This paper presents a comparison of Contention window-based MAC scheme BoP-MAC with a fragmentation based, FROG-MAC; both protocols focus on reducing the delay for higher priority traffic, while taking a diverse approach. BoP-MAC assigns a differentiated back-off value to the multi-priority traffic, whereas FROG-MAC enables early transmission of higher-priority packets by fragmenting lower-priority traffic. Simulations were performed on Contiki by varying the number of nodes for two traffic priorities. It has been shown that when working with multi-priority heterogenous data in the industrial environment, FROG-MAC results better both in terms of delay and throughput.

## Introduction

Industrial Machine-to-Machine (M2M) has been a rapidly emerging area over the past decade. Innovative applications of industrial communication have been launched such as real-time monitoring of manufacturing processes, predictive maintenance of equipment, and autonomous control systems in smart factories [1]. These applications rely on seamless communication between sensors, actuators, and controllers to ensure efficiency, reliability, and safety in industrial operations [2]. Hence, with the growing adoption of Industry 4.0 and Industrial IoT (IIoT), the demand for low-latency, high-throughput, and energy-efficient communication protocols has become increasingly critical for meeting the requirements of modern industrial systems [3, 4]. Furthermore, there exists a highly heterogenous data in the industrial environment, needing multiple levels of priority assignments, corresponding to their Quality of Service (QoS) requirements. For example, some data is categorized as higher priority (emergency shutdown signals, fault detection alerts, or real-time process control commands) as compared to other lower priority data (periodic status updates, historical logs, or maintenance reports) [5].

It is crucial to ensure that data is reliably transmitted to the required nodes within the industrial M2M networks. Moreover, it has also been one of the major challenges to ensure the timely transmission of the most critical data. Therefore, numerous MAC protocols with multi-priority capabilities have been developed to cater to diverse domains, including industrial automation [6], health monitoring [7], vehicular networks [8], and various other smart city applications [9]. In this context, contention window adaptation is an important scheme [10], where the nodes with a higher priority are given a lower value of back-off counter, leading them to a quicker channel access. However, once the lower priority data begins the transmission, the higher priority packets must wait for their complete transmission before they could begin requesting for access. In contrast, there is a packet fragmentation scheme [11], which proposed to send the lower priority packets in fragments instead of as a single unit; hence, there are some pauses of interruptible periods between each fragment, during which a higher priority packet may access the channel, resulting in reducing the delay significantly.

In this work, we compare the performance of adaptive contention scheme (by taking BoP-MAC [12] as an example) with packet fragmentation (by using an example protocol, FROG-MAC [11]) in terms of delay and throughput. The basic operation of these protocols follows in section II. We believe that it has become crucial to assess and compare the most famous MAC schemes for various applications, particularly for the critical ones such as industrial M2M communication. We conducted extensive simulations for both the protocols by varying traffic patterns, number of nodes, simulation duration etc, in order to come up with the concrete findings to support the claim that a simple fragmentation-based scheme could offer high performance in terms of delay and throughput. To demonstrate the multi-priority performance, we used low priority (normal) and high priority (urgent) traffic.

To the best of our knowledge, no previous study has compared the performance of contention window and data fragmentation-based MACs for wireless industrial networks. Hence, the major contribution of the present article is to validate the performance of a simple MAC scheme, FROG-MAC which can be used for efficiently managing industrial M2M communications.
Rest of this paper has been organized as follows: section II briefly presents the relevant work; section III details the experimental settings; section IV describes the result. Section V presents the limitations and finally, section VI concludes the paper and offers an insight into the future work directions.

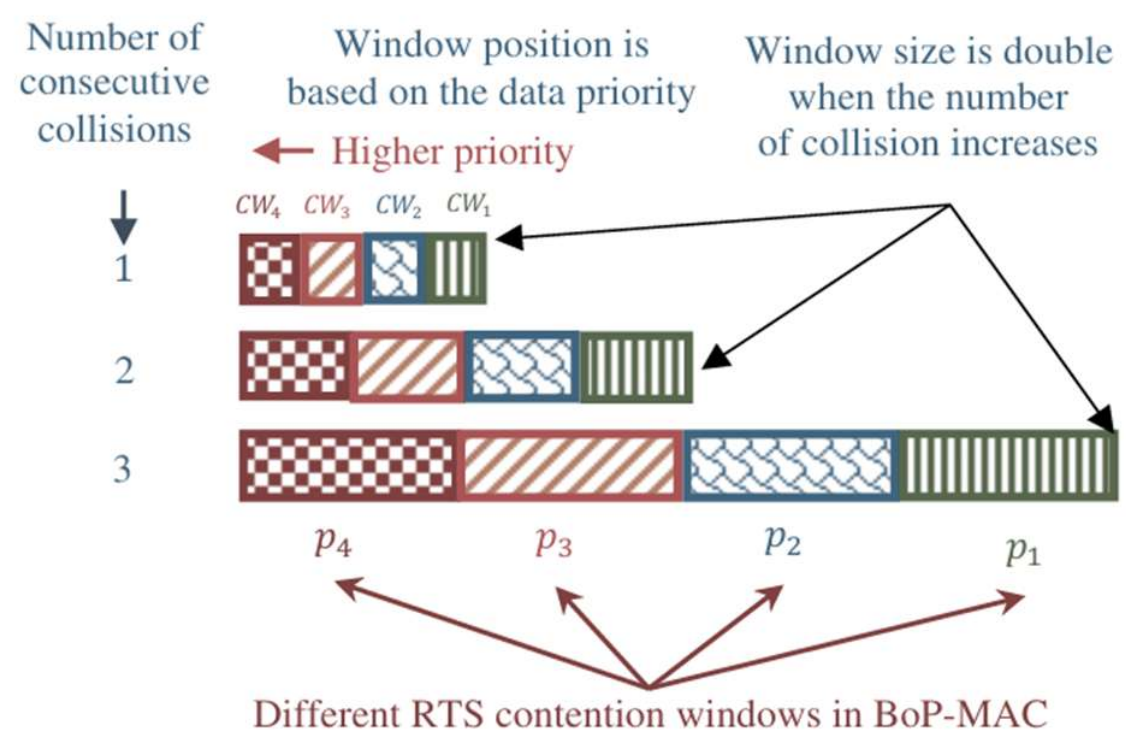

Figure 1: Adaptive Contention Based Operation of BoP-MAC

## Relevant Work

Various MAC schemes have been developed over past decade for efficiently dealing with the differentiated QoS and delay requirements of multi-priority industrial data. In this section, we briefly review some of these:
Receiver initiation and Multi-Priority Backoff (RMP-MAC) has been developed in [13], which combines the techniques of adaptive backoff and receiver initiated-wake up scheme. On one hand, the multi-priority backoff allows the protocol to run a shorter counter for urgent traffic and on the other hand, the energy efficiency is optimized using the receiver-initiated wake up mechanism. PriTraCon-MAC [14] has also been designed considering 4 traffic priorities and assigning a different contention window slot based on the traffic priority. The persistence based CSMA has also been modified in [15], where the authors attempted to adjust the p-values in accordance with the traffic priority. For all of these and similar protocols, it is evident that although it is possible to offer a better channel access to higher priority traffic, once the lower priority traffic starts transmitting, there is no possibility for urgent traffic to get hold of the channel.
A range of Machine learning (ML) based MAC protocols has also been developed to improve the transmission latency of critical data packets [17, 18]. For example,

ML centered priority-based self-organized MAC (ML-MAC) has been developed in [16], where the focus is to adjust the MAC parameters adaptively based on the ML predictions. Similarly, Deep Reinforcement Learning-Based Multi-Access (DRLMA) protocols have been proposed for dynamically switching between grant-based or grant-free channel access [19]. Although using ML to adaptively assess the traffic patterns and adjust the MAC parameters seems to be a promising area, there are some problems associated with the approach [20]; firstly, it is never a guarantee that the ML predictions efficiently model the traffic conditions, and secondly, running complex ML algorithms might not always be suitable for the end nodes. Hence, such schemes will most likely be deployed at the cluster head or edge nodes.
In this work, we selected the two most common MAC schemes, contention-window prioritization (BoP-MAC), and data fragmentation (FROG-MAC) for their comparative performance evaluation. Next, we briefly present the design principles and operation of each of these protocols.
BoP-MAC [12] mainly deploys adaptive contention window mechanism for assign-ing a differentiated channel access to normal and urgent traffic. In addition, it also uses dynamic duty-cycle and adaptive active/sleep periods. The basic operation of BoP-MAC has been shown in Figure 1, where contention window has been shown to be divided in four portions, based on the priority of each traffic class. For the higher priority traffic, the lower back-off values are assigned to ensure quicker access; however, as with the previous adaptive contention-based protocols, BoP-MAC could also not allocate the channel to urgent traffic while the normal packets were transmitting.
Finally, FROG-MAC has been described, which was developed in [11]; the basic operation of FROG-MAC has been illustrated in Figure 2. As seen in the figure, nodes with 2 traffic priorities are operating in a single-hop environment (each node directly connects with the sink). Since the focus of FROG-MAC is on providing channel access to the urgent traffic, even when the normal traffic is being transmitted, the normal traffic is always transmitted in fragments (as shown for the node 1). An interruptible period is introduced between two consecutive fragments of a normal packet, allowing a node with higher-priority traffic the opportunity to transmit a Request to Send (RTS). On the other hand, the urgent packets are not fragmented, but are sent as a single unit. While fragmenting lower-priority traffic may significantly increase its delay, it offers a substantial advantage to urgent traffic, which would otherwise have to wait for the complete transmission of normal packets. The details of differentiated acknowledgement have also been given in [11], where we reported the ways in which nodes will be acknowledging the packets received as a single unit, versus those partially received before their transmission was interrupted. To further elaborate the operation of FROG-MAC, we present the flowchart of sender's operation in Figure 3.

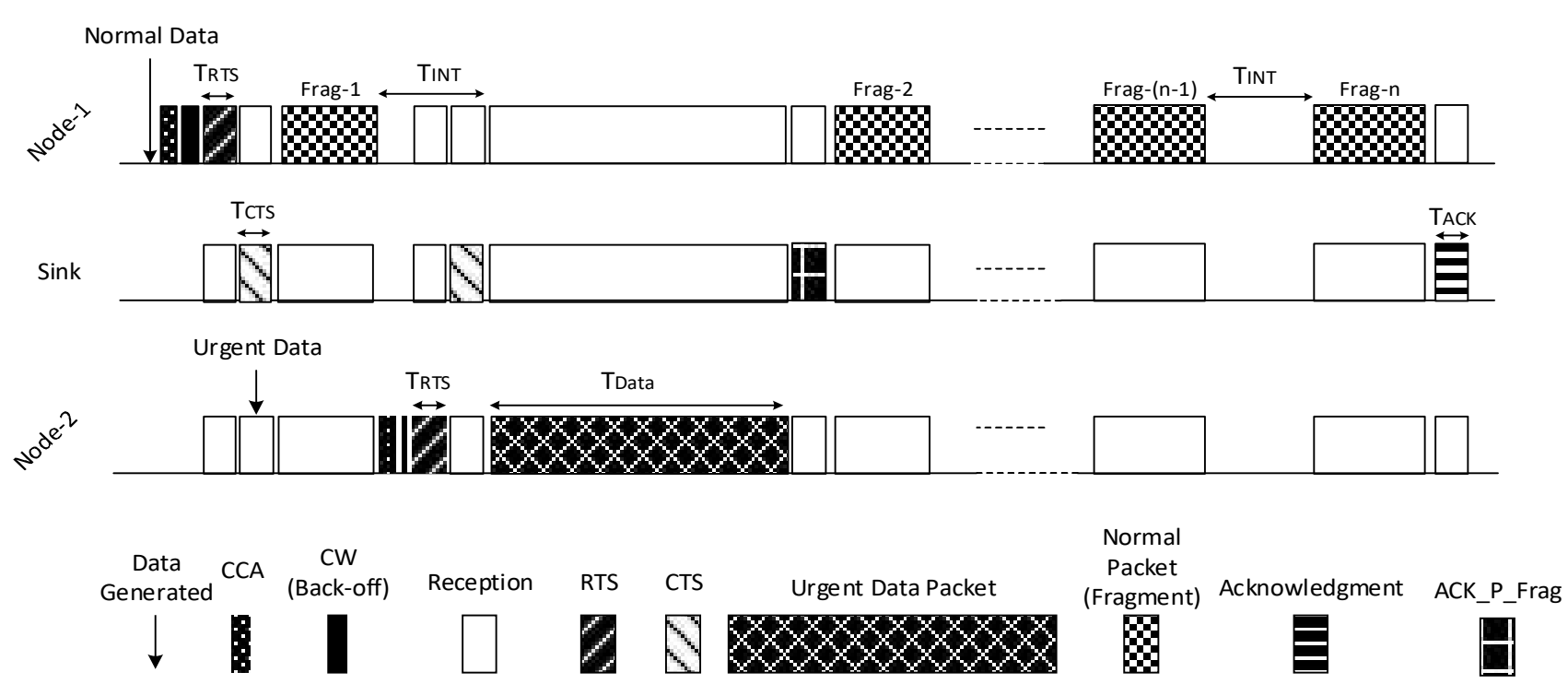

Figure 2: Basic Operation of FROG-MAC Illustrating the Use of Fragmentation of Lower Priority Data

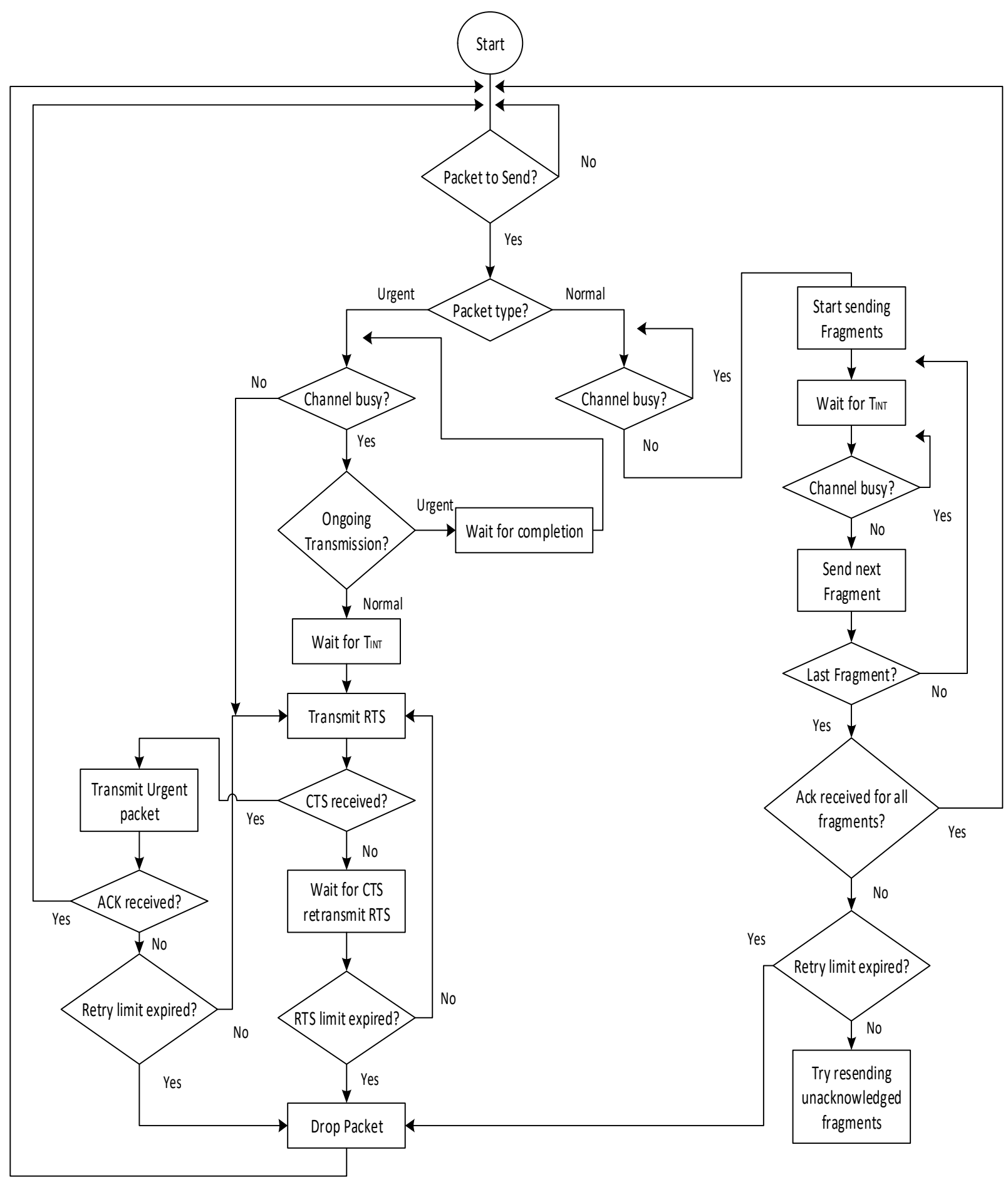

Figure 3: FROG-MAC's Sender Operation

## Experimental Settings

A single-hop star topology network comprising 11 nodes (10 source nodes and one sink) was utilized for the simulations, representing a fundamental yet practical scenario in industrial wireless networks. In this configuration, multiple source nodes communicate directly with a central sink node, such as a gateway or central controller, within their transmission range. The simulated setup is analogous to real-world industrial M2M communication, where devices transmit data, such as sensor readings or status updates, to a centralized entity for monitoring and control, enabling efficient and reliable data exchange. Additionally, two priority levels were introduced to differentiate between normal and high-urgency data. Each experiment was conducted five times, and the results were presented with a 95% confidence interval. The detailed simulation parameters are outlined in Table 1:

Table 1: Simulation Parameters

| **Simulation Parameter** | **Value** |
|---|---|
| Simulation Duration | 1000 Sec |
| Contention Window Size | Non-overlapping<br>0 to 10 for urgent traffic<br>11 to 20 for normal traffic |
| Network Size | 2 to 11 |
| Message generation interval for urgent traffic | 2 sec (Poisson) |
| Message generation interval for normal traffic | 200 msec (CBR) |
| Data Packet size | 127 Bytes |
| Total Payload length | 121 Bytes |
| Fragment size (payload length in each fragment) | Varying (2 to 121Bytes) |
| No. of Fragments | Varying (1 to 61) |
| RTS/CTS Packet size | 5 Bytes |
| ACK Packet size | 5 Bytes |
| SACK/NACK Packet size | 6 Bytes |
| ACK_P_Frag Size | 8 Bytes |
| Time taken to transmit each byte (Packetization delay) | 32 μsec |
| Interruptible Period $T_{INT}$ | 0.6 msec |

For both the protocols BoP-MAC and FROG-MAC, we used similar traffic settings. It is to be noted that Constant Bit Rate (CBR) arrivals were used for normal, whereas Poisson distribution was used to generate urgent traffic. This is in line with the nature of urgent and normal data, commonly transmitted in the industrial networks.

## Results and Discussion

We conducted performance comparisons for delay and throughput for BoP-MAC and FROG-MAC, by varying the number of nodes. The results of delay for urgent and normal traffic are illustrated in Figure 4. As previously mentioned, FROG-MAC was simulated for two fragment sizes, 16 and 2. The results for each fragment size are presented here:

As illustrated in Figure 4, delay increases with the increase in number of nodes for both traffic types, as well as for both protocols. This is because more nodes in the network implies higher congestion and channel access delays. Figure 4(a) illustrates that FROG-MAC delivers the lowest delay for urgent traffic, while exhibiting the highest delay for normal traffic. On the other hand, while BoP-MAC demonstrates better delay performance for normal traffic, its overall performance remains significantly lower than that of FROG-MAC. Hence, it has been found that BoP-MAC performs better for normal traffic, as it prioritizes access based solely on the contention window, allowing normal traffic to transmit if it wins the channel before urgent traffic. In contrast, FROG-MAC intentionally compromises the QoS for normal traffic to ensure faster transmission of urgent packets.

Similar trends are also reported in Figure 4(b). Here, the major difference has been obtained in the delays of both urgent and normal traffic for the FROG-MAC. In Figure 4(a), the fragment size was set to 16 bytes, meaning that 16 bytes of a normal packet were transmitted before the interruptible period, allowing faster transmission compared to Figure 4(b), where the fragment size was limited to 2 bytes. On the other hand, the delay for urgent traffic is further reduced compared to Figure 4(a), as these packets now have an even quicker opportunity to interrupt the channel.

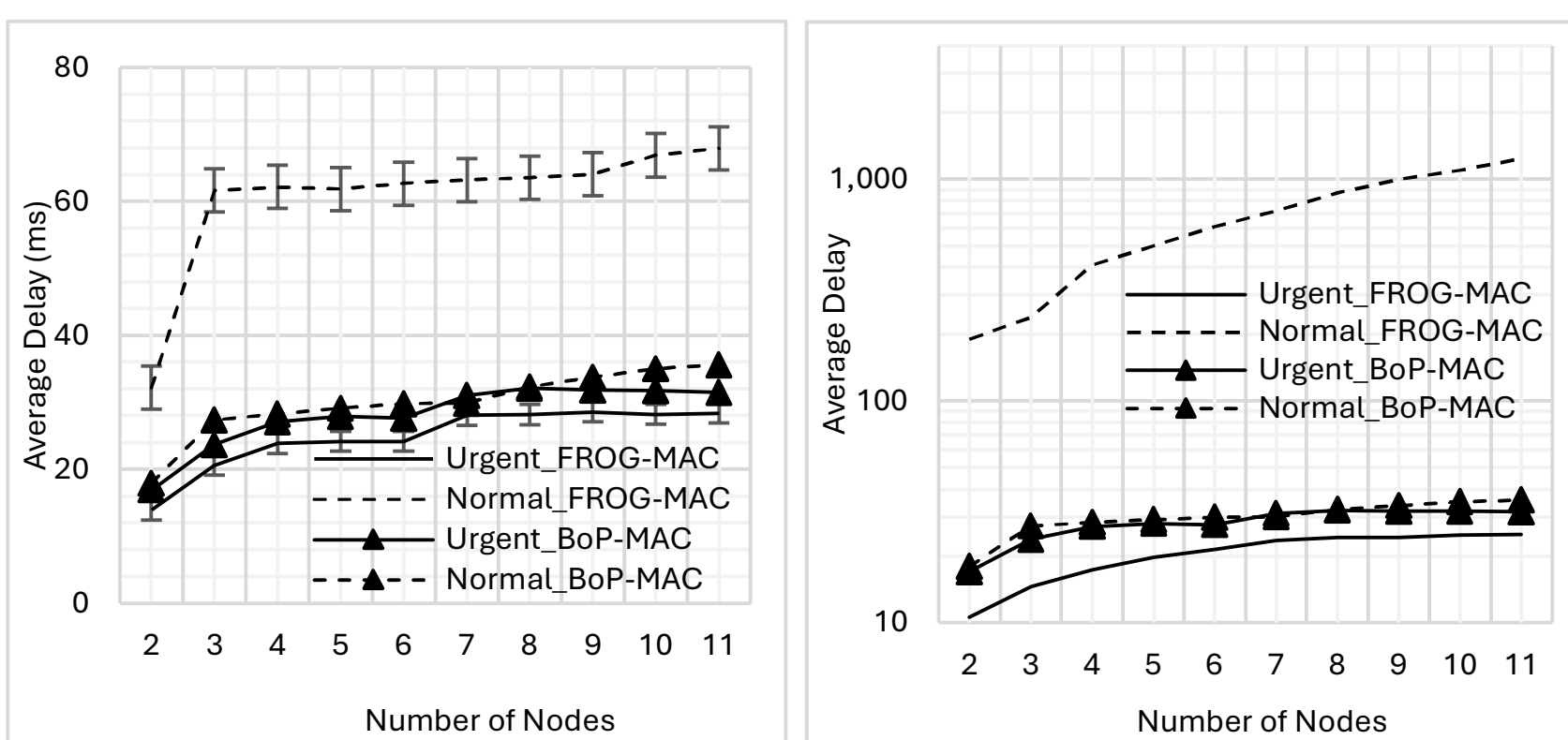

(a) (b)

Figure 4: Delay Comparison for BoP-MAC and FROG-MAC.
(a) Fragment size = 16, (b) Fragment size = 2.

The results for throughput comparison for BoP-MAC and FROG-MAC while keeping the fragment size as 16 and 2 respectively, are illustrated in Figure 5.
Figure 5(a) & (b) both illustrate the declining trend for throughput for both traffic types and protocols. This trend is observed because, as the number of nodes increases, contention for the shared medium intensifies, resulting in more frequent collisions and retransmissions. Additionally, the increased overhead from managing higher network traffic contributes to further degradation in throughput. When comparing FROG-MAC and BoP-MAC for urgent traffic in Figure 5(a), FROG-MAC remains superior even at the higher traffic loads. On the other hand, BoP-MAC performs less efficiently for urgent traffic compared to normal traffic due to the absence of fragmentation. Since it relies solely on an adaptive contention window, a normal packet that gains channel access first can occupy the medium for a longer duration, even when higher-priority traffic is waiting to transmit. Figure 5(b) subsequently compares the performance of FROG-MAC for varying the fragment size from 16 to 2. It is evident that as the fragment size decreases, the throughput of urgent traffic improves further, though at the expense of lower-priority traffic.

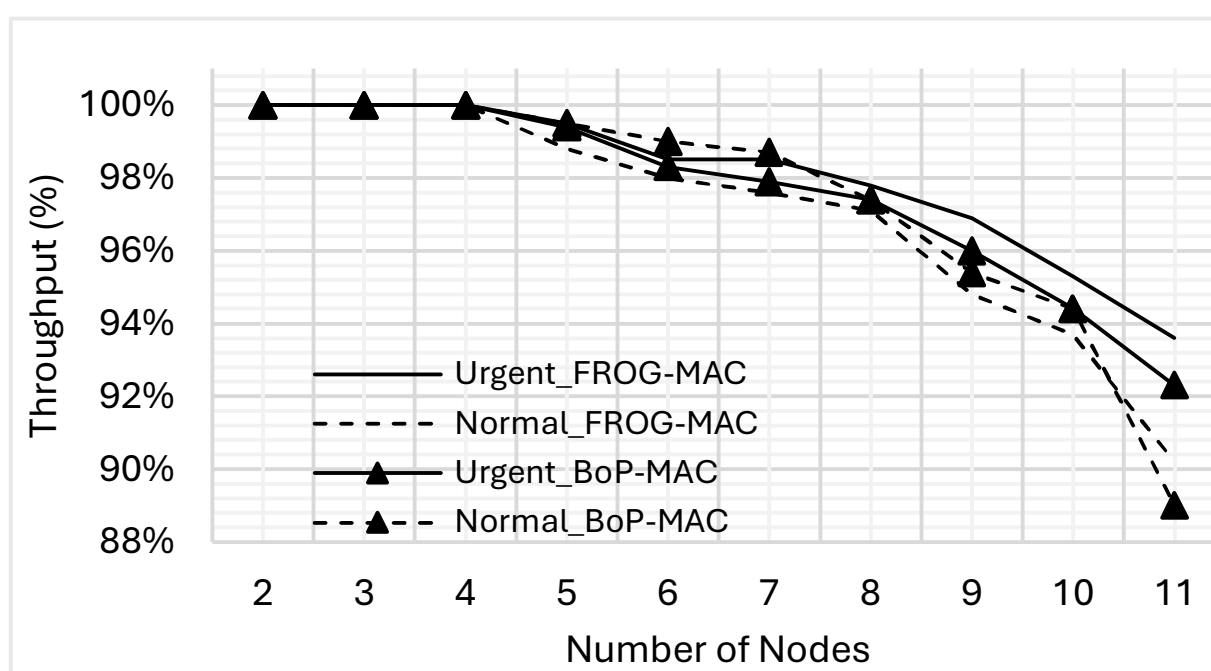

(a)

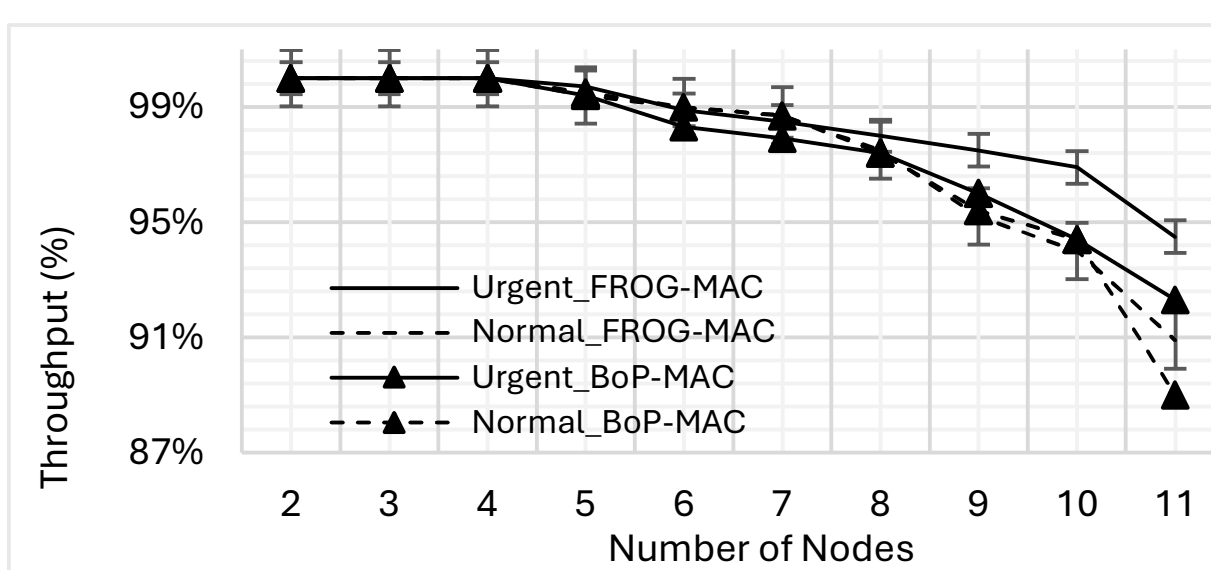

(b)

Figure 5: Throughput Comparison for BoP-MAC and FROG-MAC.
(a) Fragment size = 16, (b) Fragment size = 2.

## Study Limitations

This study has certain limitations that should be acknowledged. The performance evaluation was carried out in a controlled simulation environment with predefined traffic patterns, which may not fully represent the complexity and variability of real-world industrial settings. The analysis did not include the energy consumption aspect of FROG-MAC, which is an important factor in power-sensitive industrial applications. Furthermore, the scalability of the protocol was not evaluated for very large networks, and the protocol's responsiveness under highly dynamic traffic conditions remains an area for future exploration.

## Conclusion and Future Work

This paper presented a simulation-based comparison for the contention window and fragmentation-based MAC protocols, when operating in a multi-priority traffic based industrial M2M environments. The results show that FROG-MAC improves network performance in terms of both delay and throughput when compared to BoP-MAC. These trends have been observed due to the unique attribute of FROG-MAC which lets the higher priority data to access the channel early, by introducing pauses/interruptible periods between the transmission of lower priority data. In contrast, BoP-MAC lacks this feature and relies solely on contention window-based access, which can increase the delay for urgent traffic when lower-priority packets have already occupied the channel.

In future, we plan to improve FROG-MAC further for multi-priority data. We aim to adjust the fragment size dynamically based on the traffic arrival pattern and frequency, which is expected to enhance overall channel utilization. Moreover, emerging machine learning techniques can be integrated with FROG-MAC to better predict channel conditions and optimally determine the fragment size for each traffic priority.

## Acknowledgment

This contribution is supported by HORIZON-MSCA-2022-SE-01-01 project COALESCE under Grant Number 10113073.